\documentclass[11pt,english,aps,pra,onecolumn,tightenlines,notitlepage,floatfix,fleqn]{revtex4-2}
\renewcommand{\bibliography}[1]{}
\pdfoutput=1
\usepackage[utf8]{inputenc}
\usepackage[T1]{fontenc}
\usepackage{amsmath,amssymb,amsfonts}
\usepackage{bm,bbm}
\usepackage{epsfig}
\usepackage{graphics}

\usepackage[dvipsnames]{xcolor}
\definecolor{dblue}{rgb}{0,0.1,.6}

\usepackage[colorlinks=true,citecolor=dblue,linkcolor=dblue,urlcolor=dblue]{hyperref}
\usepackage[all]{hypcap}

\newcommand{\id}{\mathbbm{1}}
\newcommand{\bra}{\langle}
\newcommand{\ket}{\rangle}

\newcommand{\hP}{\hat{P}}

\newcommand{\s}{\sigma}
\newcommand{\CC}{\mathbb{C}}

\newcommand{\mc}[1]{\mathcal{#1}}
\newcommand{\pdag}{{\phantom{\dag}}}

\renewcommand{\H}{\mc{H}}
\newcommand{\E}{\mc{E}}
\renewcommand{\L}{\mc{L}}
\newcommand{\R}{\mc{R}}
\newcommand{\veps}{\varepsilon}
\newcommand{\rhot}{\tilde{\rho}}
\newcommand{\SVD}{\text{SVD}}
\newcommand{\trunc}{\text{trunc}}
\newcommand{\TTNS}{\text{TTNS}}

\newcommand{\Emph}[1]{\emph{\textbf{#1}}}

\makeatletter
\def\@hangfrom@section#1#2#3{\@hangfrom{#1#2}#3}
\def\@hangfroms@section#1#2{#1#2}
\makeatother

\newcommand{\qlab}  {National Quantum Laboratory, University of Maryland, College Park, MD 20742, USA}
\newcommand{\umd}   {Department of Physics, University of Maryland, College Park, MD 20742, USA}
\newcommand{\duke}  {Department of Physics and Duke Quantum Center, Duke University, Durham, NC 27708, USA}

\begin{document}

\title{Tree tensor network states represent low-energy states faithfully}
\author{Thomas Barthel}
\affiliation{\qlab\vspace{-1em}}
\affiliation{\umd\vspace{-1em}}
\affiliation{\duke}
\date{December 24, 2025}

\begin{abstract}
Extending corresponding results for matrix product states [Verstraete and Cirac, PRB \textbf{73}, 094423 (2006); Schuch et al.\ PRL \textbf{100}, 030504 (2008)], it is shown how the approximation error of tree tensor network states (TTNS) can be bounded using Schmidt spectra or R\'{e}nyi entanglement entropies of the target quantum state. Conversely, one obtains bounds on TTNS bond dimensions needed to achieve a specific approximation accuracy. The result implies that efficient TTNS approximations exist if $\alpha<1$ R\'{e}nyi entanglement entropies for single-branch cuts obey an area law, as in ground and low-energy states of certain gapped systems on tree graphs.
\end{abstract}
\vspace{-0.5em}

\maketitle

\section{Introduction and main result}\label{sec:Intro}
\vspace{-0.5em}
\emph{Tensor network states} (TNS) \cite{Kramers1941-60,Baxter1968-9,Nightingale1986-33,Fannes1992-144,White1992-11,Niggemann1997-104,Nishino2000-575,Verstraete2004-7,Vidal-2005-12,Schollwoeck2011-326,Orus2014-349,Cirac2021-65} are an important tool for the investigation of strongly correlated quantum many-body systems. Numerical TNS algorithms allow for the study of ground states, finite-temperature states, response functions, non-equilibrium dynamics, driven-dissipative systems, and quantum circuits. The basic idea is to approximate many-body states by a network of partially contracted tensors. The tensors are assigned to vertices of a graph and can carry physical indices. In addition, neighboring tensors share common bond indices that we sum over to obtain the quantum state. When the network structure of the TNS aligns well with the entanglement structure of the target state, small bond dimensions (small numbers of effective degrees of freedom) suffice for accurate TNS approximations, thus bypassing the curse of dimensionality.

Following Steve White's development of the \emph{density matrix renormalization group} (DMRG) for one-dimensional systems \cite{White1992-11,White1993-10}, it was realized \cite{Oestlund1995,Rommer1997} that DMRG is actually a variational algorithm for \emph{matrix product states} (MPS) \cite{Baxter1968-9,Affleck1987-59,Fannes1992-144,Rommer1997}.
Around the same time, the DMRG idea was adapted for the simulation of quantum systems on tree graphs \cite{Otsuka1996-53,Friedman1997-9,Lepetit2000-13},
and it turns out that this yields a variational algorithm for \emph{tree tensor network states} (TTNS), which have precursors in real-space renormalization group schemes \cite{Kadanoff1966-2,Jullien1977-38,Drell1977-16,Martin-Delgado1996-76,Fisher1998-70} and had previously been considered in analytical work \cite{Affleck1988-115,Fannes1992-66}.
Use and method development accelerated after TTNS algorithms were formulated in the language of tensor networks \cite{Shi2006-74,Tagliacozzo2009-80,Murg2010-82,Nakatani2013-138}.
Both being loop-free classes of tensor networks, the MPS and TTNS varieties are closed sets \cite{Barthel2022-112} and the variational optimization is free of barren plateaus \cite{Barthel2023_03,Miao2024-109}.

As previously mentioned, the TNS tensors can carry physical indices like $\s_i=\uparrow,\downarrow$ that label basis states of local quantum degrees of freedom like a spin-1/2. In addition, all tensors carry the auxiliary \emph{bond indices} $\mu_i=1,\dotsc,M_i$. The computational efficiency critically depends on the required bond dimensions $M_i$ needed to achieve a specific approximation accuracy. For MPS, Refs.~\cite{Verstraete2005-5,Schuch2008-100a} established upper and lower bounds on bond dimensions, using Schmidt spectra or R\'{e}nyi entanglement entropies of the target quantum state. Given that TTNS have found many applications for the simulation of strongly correlated systems spanning condensed matter \cite{Tagliacozzo2009-80,Murg2010-82,Bauernfeind2017-7,Gerster2017-96,Macaluso2020-2,Kloss2020-9,Bauernfeind2020-8,Cao2021-104,Cataldi2021-5,Felser2021-126,Hikihara2023-5,Grundner2024-109,Sulz2024-109,Pavesic2025-111,Krinitsin2025-112,Krinitsin2025-134,Pavesic2025_09}, nuclear and particle physics \cite{Felser2020-10,Magnifico2021-12,Cataldi2025_09}, quantum chemistry \cite{Murg2010-82,Nakatani2013-138,Murg2015-11,Gunst2018-14,Larsson2019-151}, and quantum computation \cite{Dumitrescu2017-96,Seitz2023-7,Jaschke2024-9,Dubey2025-112}, it is important to devise corresponding bounds for TTNS. Note that TTNS are successfully applied to both systems with \cite{Otsuka1996-53,Friedman1997-9,Lepetit2000-13,Shi2006-74,Nakatani2013-138,Bauernfeind2017-7,Bauernfeind2020-8,Cao2021-104,Grundner2024-109} and without \cite{Tagliacozzo2009-80,Murg2010-82,Nakatani2013-138,Gerster2017-96,Macaluso2020-2,Kloss2020-9,Cataldi2021-5,Felser2021-126,Hikihara2023-5,Sulz2024-109,Pavesic2025-111,Krinitsin2025-112,Krinitsin2025-134,Pavesic2025_09,Felser2020-10,Magnifico2021-12,Cataldi2025_09,  Murg2015-11,Gunst2018-14,Larsson2019-151,Dumitrescu2017-96,Seitz2023-7,Jaschke2024-9,Dubey2025-112,Tindall2024_10} physical tree structure.

This work establishes lower and upper bounds on TTNS approximation errors and required TTNS bond dimensions based on Schmidt spectra and R\'{e}nyi entanglement entropies \cite{Renyi1961,Laflorencie2016-646} of the target state.

\Emph{Main result.}\,---
\emph{Consider a tree graph with $N$ vertices, an assignment of quantum degrees of freedom to vertices of the graph, and a normalized target state $|\psi\ket$ of the composite system. Let $\lambda_{i,1}\geq \lambda_{i,2}\geq\dotsc\geq 0$ denote the Schmidt coefficients of $|\psi\ket$ for the bipartition of the system that corresponds to cutting edge $i$, with $S_{\alpha,i}=\frac{1}{1-\alpha}\ln\sum_\mu\lambda_{i,\mu}^{2\alpha}$ being the associated $\alpha$-R\'{e}nyi entanglement entropy.}

\emph{\textbf{(a)}\ There exists a TTNS approximation $|\psi_\TTNS\ket$ with bond dimensions $M_1,\dotsc,M_{N-1}$ that achieves an approximation accuracy $\delta:=\|\psi-\psi_\TTNS\|^2$ bounded by
\begin{gather}\label{eq:MR-errTot}
	1-\min_i\left(\frac{M_i}{e^{S_{\tilde{\alpha},i}}}\right)^{1-\frac{1}{\tilde{\alpha}}}
	\leq \max_i \veps_i(M_i)\,\, \leq\,\, \delta\,\, \leq\,\, \sum_i \veps_i(M_i)
	\leq \sum_i \left(\frac{e^{S_{\alpha,i}}}{M_i}\right)^{\frac{1}{\alpha}-1},\\
	\label{eq:MR-err}
	\text{where}\quad
	\veps_i(M_i):=\sum_{\mu>M_i}\lambda^2_{i,\mu}
\end{gather}
is the bond-$i$ truncation error, and the R\'{e}nyi parameters can be freely chosen in the ranges $\tilde{\alpha}>1$ and $\veps_i\leq\alpha<1$, respectively.}

\emph{\textbf{(b)}\ For every uniform bound $\epsilon>0$ on the single-bond truncation errors, there exists a TTNS approximation $|\psi_\TTNS\ket$ with approximation accuracy $\delta:=\|\psi-\psi_\TTNS\|^2\leq (N-1)\,\epsilon$ and bond dimensions $M_1,\dotsc,M_{N-1}$ bounded by
\begin{equation}\label{eq:MR-Mbound}
	e^{S_{\tilde{\alpha},i}}(1-\delta)^{\frac{\tilde{\alpha}}{\tilde{\alpha}-1}}
	\leq \min\{ M \,|\, \veps_i(M)\leq\delta\}
	\leq M_i\leq \min\{ M \,|\, \veps_i(M)\leq\epsilon\}
	\leq e^{S_{\alpha,i}} \left(\frac{1}{\epsilon}\right)^{\frac{\alpha}{1-\alpha}}
\end{equation}
for any $\tilde{\alpha}>1$ and $\alpha$ with $\epsilon\leq\alpha<1$.}
\vspace{0.7em}

Section~\ref{sec:Notations} introduces notations and conventions. Section~\ref{sec:ExactTTNS} describes how to construct exact TTNS representations for any target state $|\psi\ket$ by a sequence of singular value decompositions (SVD). Section~\ref{sec:Lower} explains the lower bounds in Eqs.~\eqref{eq:MR-errTot} and \eqref{eq:MR-Mbound} in terms of truncation errors. Section~\ref{sec:truncTTNS-simple} briefly discusses a simple and instructive TTNS approximation scheme that leads to an upper bound on the approximation accuracy similar to Eq.~\eqref{eq:MR-errTot}, with the caveat that the bound depends on Schmidt spectra of truncated states instead of those of the target state $|\psi\ket$. Starting from the exact TTNS representation, Section~\ref{sec:truncTTNS-proper} proves the upper bounds in the main result in terms of truncation errors. Finally, Section~\ref{sec:Renyi} explains the bounds in terms of R\'{e}nyi entanglement entropies.
Section~\ref{sec:Conclusion} provides a conclusion and comments on entanglement area laws for quantum systems on tree graphs.

\section{Notations and conventions}\label{sec:Notations}
Consider a quantum many-body system with $N$ sites. Each site $i$ is associated with a Hilbert space of dimension $d_i$ and orthonormal single-site basis states
\begin{equation}
	|\s_i\ket\quad\text{with}\quad \s_i=1,\dotsc,d_i.
\end{equation}
For distinguishable $d_i$-level systems, the total Hilbert space $\H$ is the tensor product space $\bigotimes_{i=1}^N \CC^{d_i}$. For identical fermions, we are restricted to the anti-symmetric subspace with Fock-state basis $\{|\s_1,\dotsc,\s_N\ket\}$ \cite{Barthel2009-80,Corboz2009-80,Corboz2009_04,Pineda2009_05,Kraus2009_04}. For bosons, we may limit maximum occupation numbers in accordance with energy constraints.

For corresponding TTNS, the physical sites are assigned to the vertices of a tree graph with vertex $i=N$ (arbitrarily) chosen as the root and vertex labels decreasing with increasing graph distance from the root. An example is shown in Fig.~\ref{fig:ExactTTNS}a. Edges of the graph are directed from parents to children, where edge $i$ is pointing towards vertex $i$. Cutting edge $i$ decomposes the tree graph into the branch $\L_i$, containing vertex $i$ and its descendants, and the remainder $\R_i=[1,N]\setminus \L_i$.

We assign bond vector spaces of dimension $M_i$ to each edge and assign one tensor $A_i$ to each vertex. If vertex $i$ has $y$ children $j_1,\dotsc,j_y$, then $A_i$ can be considered a linear map
\begin{equation}\label{eq:A}
	A_i: \CC^{M_i} \to \CC^{d_i}\otimes\CC^{M_{j_1}}\otimes\dotsb\otimes\CC^{M_{j_y}}
\end{equation}
from the bond-$i$ vector space to the tensor product of the physical site-$i$ vector space and the child bond vector spaces.
The tensor elements can be labeled as
\begin{equation}\label{eq:A-elem}
	[A_i^{\s_i}]_{(\mu_{j_1},\dotsc,\mu_{j_y}),\mu_i}\quad\text{with}\quad
	\s_i=1,\dotsc,d_i,\quad \text{and}\quad
	\mu_j=1,\dotsc,M_j.
\end{equation}
For the root tensor $A_N$, we can set $M_N=1$ and interpret $A_N$ as a pure quantum state (vector) on which the linear maps $A_{N-1},...,A_1$ act to generate the TTNS in the Hilbert space $\H$. Equivalently, the TTNS in the basis $\{|\s_1,\dotsc,\s_N\ket\}$ is the product of the tensor elements \eqref{eq:A-elem}, summed  over all bond indices $\mu_1,\dotsc,\mu_{N-1}$. For a TTNS with ``internal'' tensors $A_i$ that do not carry a physical index, we can set the corresponding site-space dimensions to $d_i=1$.

We say that the TTNS is in canonical form with vertex $N$ as the orthogonality center \cite{Schollwoeck2011-326}, if all tensors $A_i$ are partial isometries and their adjoints $A_i^\dag$ are (injective) isometries, i.e., if we have the orthonormality conditions
\begin{equation}\label{eq:A-ON}
	A_i^\dag A^\pdag_i \equiv \sum_\s A_i^{\s\dag} A_i^{\s\pdag} = \id_{M_i}.
\end{equation}

\section{Exact TTNS representation}\label{sec:ExactTTNS}
We want to obtain an exact canonical TTNS representation of a normalized quantum state
\begin{equation}\label{eq:psi}
	|\psi\ket = \sum_{\s_1,\dotsc,\s_N} \Psi^{\s_1,\dotsc,\s_N}|\s_1,\dotsc,\s_N\ket.
\end{equation}

This can be achieved by a sequence of $i=1,\dotsc,N-1$ SVDs, each corresponding to a Schmidt decomposition
\begin{equation}\label{eq:exact-Schmidt}
	|\psi\ket = \sum_{\mu} \lambda_{i,\mu}|\ell_{i,\mu}\ket\otimes|r_{i,\mu}\ket\quad\text{with Schmidt coefficients}\quad
	1\geq \lambda_{i,1}\geq\lambda_{i,2}\geq \dots\geq 0
\end{equation}
and orthonormal bases
\begin{equation}
	\{|\ell_{i,\mu}\ket\}_\mu\quad\text{and}\quad
	\{|r_{i,\mu}\ket\}_\mu
\end{equation}
for the subsystems $\L_i$ and $\R_i$, respectively, corresponding to a cut of edge $i$ as defined in Section~\ref{sec:Notations}.

\begin{figure}[t]
	\includegraphics[width=0.92\textwidth]{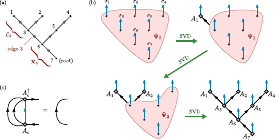}
	\caption{\label{fig:ExactTTNS}
	For an exact TTNS representation of a target state \eqref{eq:psi} with expansion coefficients $\Psi^{\s_1,\dotsc,\s_N}$,
	(a) assign the physical sites to vertices of a directed tree graph, where cutting edge $i$ yields a bipartition into subsystem $\L_i$ and its complement $\R_i$.
	(b) A sequence of $N-1$ SVDs \eqref{eq:SVD} or, equivalently, QR decompositions then splits off one TTNS tensor $A_i$ in each step.
	(c) The resulting tensors are isometries with the orthonormality property \eqref{eq:A-ON}.}
\end{figure}

\begin{figure}[t]
	\includegraphics[width=0.77\textwidth]{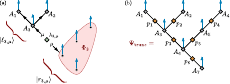}
	\caption{\label{fig:Schmidt-trunc}
	(a) The SVD \eqref{eq:SVD} for edge $i$ yields a Schmidt decomposition \eqref{eq:exact-Schmidt} of the target state $|\psi\ket$ with the TTNS formed by tensors $\{A_j\,|\,j\in\L_i\}$ providing an orthonormal basis $\{|\ell_{i,\mu}\ket\}$ for subsystem $\L_i$ -- in this case, sites 1, 2 and 3. Similarly, tensor $\Phi_i$ provides an orthonormal basis $\{|r_{i,\mu}\ket\}$ for subsystem $\R_i$ -- in this case, sites 4, 5, 6 and 7. 
	(b) The truncated TTNS \eqref{eq:psiTrunc} is obtained from the exact TTNS representation of the target state by inserting projection operators $p_i$ at every edge, only retaining the $M_i$-dimensional bond subspace corresponding to the largest Schmidt coefficients $\lambda_{i,1}\geq\dotsc\geq\lambda_{i,M_i}$.}
\end{figure}

On the level of the expansion coefficients $\Psi=:\Psi_0$ and TTNS tensors, this means
\begin{equation}\label{eq:SVD}
	\Psi_{i-1}^{\s_i,\dotsc,\s_N}\stackrel{\SVD}=: A_i^{\s_i}\lambda_i \Phi_i^{\s_{i+1},\dotsc,\s_N}
	\quad\text{and we define}\quad
	\Psi_{i-1}^{\s_{i+1},\dotsc,\s_N}:=\lambda_i \Phi_i^{\s_{i+1},\dotsc,\s_N}
\end{equation}
for the next SVD as visualized in Figs.~\ref{fig:ExactTTNS}b and \ref{fig:Schmidt-trunc}.
Here, we only wrote physical indices but no bond indices, $A_i$ obeys the orthonormality condition \eqref{eq:A-ON}, and $\lambda_i$ is the diagonal matrix of Schmidt coefficients on the bond vector space of edge $i$. The tensor network formed by the tensors $\{A_j\,|\,j\in\L_i\}$ with fixed index $\mu$ on bond $i$ gives the state $|\ell_{i,\mu}\ket$ from Eq.~\eqref{eq:exact-Schmidt}, and $\Phi_i$ is the tensor of expansion coefficients for the states $|r_{i,\mu}\ket$. See Fig.~\ref{fig:Schmidt-trunc}a. Orthonormality $\bra\ell_{i,\mu}|\ell_{i,\mu'}\ket=\delta_{\mu,\mu'}$ and $\bra r_{i,\mu}|r_{i,\mu'}\ket=\delta_{\mu,\mu'}$ of these states follows from Eq.~\eqref{eq:A-ON} and an analog orthonormality property of the tensor $\Phi_i$, respectively.

\section{Lower bounds on error and bond dimensions of truncated TTNS}\label{sec:Lower}
The lower bound $\max_i \veps_i(M_i)$ for the approximation accuracy $\delta:=\|\psi-\psi_\TTNS\|^2$ in Eq.~\eqref{eq:MR-errTot} follows from the Eckart–Young–Mirsky theorem \cite{Eckart1936-1}: The best rank-$M_i$ approximation to the target state $|\psi\ket$ is given by retaining the $M_i$ largest Schmidt components of the decomposition \eqref{eq:exact-Schmidt}. Hence, $\delta$ is bounded from below by each of the rank-$M_i$ truncation errors \eqref{eq:MR-err}.

Similarly, in order to obtain a TTNS approximation with accuracy $\delta$, the bond dimension for edge $i$ cannot be so small that the single-edge truncation error already exceeds $\delta$, i.e., $M_i\geq \min\{ M \,|\, \veps_i(M)\leq\delta\}$. This is the lower bound in Eq.~\eqref{eq:MR-Mbound}.

\section{Lazy upper bound on error of truncated TTNS}\label{sec:truncTTNS-simple}
In general, bond dimensions for the exact TTNS representation of Section~\ref{sec:ExactTTNS} grow exponentially in the system size $N$. We are interested in efficient TTNS approximations and want to determine the upper bounds on the error introduced by truncating bond dimensions to some given values $M_1,\dotsc,M_{N-1}$ or, conversely, want to determine useful upper bounds on bond dimensions $M_i$ that allow for TTNS approximations of a certain accuracy.

One approach is to simply truncate the bond vector space of edge $i$ in each of the Schmidt decompositions \eqref{eq:exact-Schmidt} to dimension $M_i$, i.e., only retain the $M_i$ largest singular values before proceeding to the decomposition for edge $i+1$. In this way, we generate a sequence of truncated states
\begin{equation}\label{eq:psi_i-simple}
	|\psi\ket\equiv |\psi_0\ket \mapsto |\psi_1\ket \mapsto |\psi_2\ket \mapsto \dots \mapsto |\psi_{N-1}\ket,
\end{equation}
where $|\psi_{N-1}\ket$ is a TTNS with bond dimensions $\{M_i\}$.
Each truncation introduces an error
\begin{equation}
	\veps'_i(M_i):=\|\psi_i-\psi_{i-1}\|^2=\sum_{\mu>M_i}(\lambda'_{i,\mu})^2.
\end{equation}
Here, primes indicate that the Schmidt coefficients $\lambda'_{i,\mu}$ of state $|\psi_{i-1}\ket$ generally deviate from the exact Schmidt coefficients $\lambda_{i,\mu}$ of the target state $|\psi\ket$ in Eq.~\eqref{eq:exact-Schmidt}.

The approximation accuracy can be determined using the telescoping sum
\begin{equation}
	|\psi\ket-|\psi_{N-1}\ket
	= \underbrace{|\psi_0\ket-|\psi_1\ket}_{=:|\Delta'_1\ket}
	+ \underbrace{|\psi_1\ket-|\psi_2\ket}_{=:|\Delta'_2\ket} + \dotsc
	+ \underbrace{|\psi_{N-2}\ket-|\psi_{N-1}\ket}_{=:|\Delta'_{N-1}\ket}.
\end{equation}
Now, $|\Delta'_i\ket$ is orthogonal to $|\Delta'_j\ket$ for all $j>i$ because $|\psi_i\ket$ and all its successors have no support on the corresponding discarded subspace for subsystem $\L_i$. Hence, repeated application of the Pythagorean theorem yields the approximation accuracy
\begin{equation}\label{eq:errTot-simple}
	\|\psi-\psi_{N-1}\|^2 =  \|\Delta'_1\|^2 + \dots + \|\Delta'_{N-1}\|^2 = \sum_{i=1}^{N-1} \veps'_i(M_i).
\end{equation}

\section{Proper upper bound on error of truncated TTNS}\label{sec:truncTTNS-proper}
While Eq.~\eqref{eq:errTot-simple} nicely connects the approximation accuracy to Schmidt spectra of the states $|\psi_i\ket$ in Eq.~\eqref{eq:psi_i-simple} and is similar to the upper bound in Eq.~\eqref{eq:MR-errTot}, there is one important caveat: We want to bound TTNS approximation errors and necessary bond dimensions using the Schmidt spectra and R\'{e}nyi entanglement entropies of the target state $|\psi\ket$ itself. Of course, the Schmidt spectra and entanglement entropies of the truncated states $|\psi_i\ket$ should be similar to those of the target state $|\psi\ket$ if truncation errors are chosen small enough. However, tight continuity bounds for R\'{e}nyi entropies due to Audenaert involve the Hilbert space dimension \cite{Audenaert2007-40} (see also Refs.~\cite{Hanson2017_07,Rastegin2011-143,Chen2017-60}), reducing their value and the practicability of Eq.~\eqref{eq:errTot-simple} for our purposes.

\Emph{MPS.}\,---
In Ref.~\cite{Verstraete2005-5}, Verstraete and Cirac established an approximation accuracy bound for MPS based on the Schmidt spectra $\lambda_i$ of the target state $|\psi\ket$: An exact MPS representation $\Psi^{\s_1,\dotsc,\s_N}=A_1^{\s_1}A_2^{\s_2}\dots A_N^{\s_N}$ can be constructed as described in Section~\ref{sec:ExactTTNS} for the special case of a linear graph (every vertex $i>1$ having exactly one child $i-1$ such that $\L_i=[1,i]$ and $\R_i=[i+1,N]$). Inserting at every bond $i$ a projection
\begin{equation}\label{eq:p}
	[p_i]_{\mu,\mu'}=
	\begin{cases}
	\delta_{\mu,\mu'}&\text{if}\ \mu\leq M_i,\\
	0 & \text{otherwise}\end{cases}
\end{equation}
onto the subspace corresponding to the $M_i$ largest Schmidt coefficients $\lambda_{i,\mu}$, we obtain the truncated matrix product $\Psi_\trunc^{\s_1,\dotsc,\s_N}=\tilde{A}_1^{\s_1}\tilde{A}_2^{\s_2}\dots \tilde{A}_N^{\s_N}$, where $\tilde{A}_i^\s:=p_{i-1}A_i^\s p_{i}$ is the upper-left $M_{i-1}\times M_i$ block of the matrix $A_i^\s$. Verstraete and Cirac wrote the overlap of the target state $|\psi\ket$ and the truncated MPS $|\psi_\trunc\ket$ with a sequence of quantum channels
\begin{equation}
	\bra\psi|\psi_\trunc\ket = \Psi^\dag\Psi_\trunc = \E_1(\E_2(\dotsc\E_{N-2}(\E_{N-1}(\lambda^2_{N-1} p_{N-1})p_{N-2})\dotsc p_2)p_1),
\end{equation}
where the quantum channels
\begin{equation}\label{eq:MPS-E}
	\E_i(\rho):=\sum_\s A_i^{\s\pdag}\! \rho\, A_i^{\s\dag}
	\quad\text{have the property}\quad
	\E_i(\lambda^2_i)=\lambda^2_{i-1}
\end{equation}
due to Eq.~\eqref{eq:SVD}. Using this property, a short calculation yields the MPS approximation accuracy
\begin{equation}\label{eq:MPS-VCbound}
	\|\psi-\psi_\trunc\|^2\leq 2\sum_i\sum_{\mu>M_i}\lambda^2_{i,\mu}.
\end{equation}
See also the appendix of Ref.~\cite{Barthel2017_08}.

\Emph{TTNS.}\,---
This approach does not work for TTNS that comprise vertices $i$ with multiple children because a corresponding quantum channel $\E_i$, mapping into the tensor product of the children's bond vector spaces, does not have the property from Eq.~\eqref{eq:MPS-E}. Instead, we can proceed as follows.

Starting from the exact TTNS representation of the target state $|\psi\ket$ constructed in Section~\ref{sec:ExactTTNS} and the Schmidt decompositions \eqref{eq:exact-Schmidt}, we can define one projection operator
\begin{equation}
	\hP_i:=\sum_{\mu\leq M_i}|\ell_{i,\mu}\ket\bra\ell_{i,\mu}|\otimes \id_{\R_i}
\end{equation}
for every edge $i=1,\dotsc,N-1$ of the tree graph. $\hP_i$ acts non-trivially on subsystem $\L_i$ (the branch of vertex $i$) and as the identity on subsystem $\R_i$ (the remainder). Specifically,
\begin{equation}
	\hP_i|\psi\ket \stackrel{\eqref{eq:exact-Schmidt}}{=} 
	\sum_{\mu\leq M_i} \lambda_{i,\mu}|\ell_{i,\mu}\ket\otimes|r_{i,\mu}\ket
\end{equation}
is the original exact TTNS with the sum over bond index $\mu_i$ restricted to the range $[1,M_i]$, corresponding to the $M_i$ largest Schmidt coefficients $\lambda_{i,1},\dotsc,\lambda_{i,M_i}$. Equivalently, $\hP_i|\psi\ket$ is obtained by inserting the projector $p_i$ from Eq.~\eqref{eq:p} on edge $i$ of the exact TTNS representation. The associated truncation error is
\begin{equation}\label{eq:err}
	\big\|(1-\hP_i)|\psi\ket\big\|^2 = \sum_{\mu>M_i}\lambda^2_{i,\mu} \stackrel{\eqref{eq:MR-err}}{\equiv}\veps_i(M_i) .
\end{equation}

The (globally) truncated TTNS, as illustrated in Fig.~\ref{fig:Schmidt-trunc}b, is obtained from the exact representation by restricting all bond indices $\mu_1,\mu_2,\dotsc,\mu_{N-1}$ \emph{simultaneously} to the ranges $\mu_i=1,\dotsc,M_i$. It can be written in the form
\begin{equation}\label{eq:psiTrunc}
	|\psi_\trunc\ket = \hP_1\hP_2\dotsb\hP_{N-1}|\psi\ket.
\end{equation}
The projectors $\{\hP_i\}$ generally do not commute, and the sequence is essential for the equivalence of Eq.~\eqref{eq:psiTrunc} to the simultaneous insertion of bond-space projectors $p_i$ in the TTNS. In this sequence, projector $\hP_i$ does not influence the tensors of vertices $j<i$; hence, a subsequent action of a projector $\hP_{j<i}$ still corresponds exactly to the insertion of $p_j$ on edge $j$ of the TTNS.

Using a telescoping sum, the difference from the target state can be written as
\begin{align}
	|\psi\ket-|\psi_\trunc\ket&
	=(1-\hP_1\hP_2\dotsb\hP_{N-1})|\psi\ket \nonumber\\
	&=(1-\hP_1)|\psi\ket+\hP_1(1-\hP_2\hP_3\dotsb\hP_{N-1})|\psi\ket = \dots \nonumber\\
	&=\underbrace{(1-\hP_1)|\psi\ket}_{=:|\Delta_1\ket} + \underbrace{\hP_1(1-\hP_2)|\psi\ket}_{=:|\Delta_2\ket} + \dotsc + \underbrace{\hP_1\hP_2\dotsb\hP_{N-2}(1-\hP_{N-1})|\psi\ket}_{=:|\Delta_{N-1}\ket}.
	\label{eq:errDelta}
\end{align}
The deviation $|\Delta_1\ket$ is orthogonal to all $|\Delta_{j>1}\ket$. Due to the non-commuting projectors, $|\Delta_{i>1}\ket$ is generally \emph{not} orthogonal to all $|\Delta_j\ket$ with $j>i$. However, we can still bound the approximation accuracy through repeated application of the Pythagorean theorem $\big\|(1-\hP)|\psi\ket+\hP|\phi\ket\big\|^2=\big\|(1-\hP)|\psi\ket\big\|^2+\big\|\hP|\phi\ket\big\|^2$ and the contractive property $\big\|\hP|\phi\ket\big\|\leq\big\||\phi\ket\big\|$ of orthogonal projections. With the definition
\begin{equation}
	|\phi_i\ket:=(1-\hP_i)|\psi\ket + \hP_i(1-\hP_{i+1})|\psi\ket + \dotsc + \hP_i\hP_{i+1}\dotsb\hP_{N-2}(1-\hP_{N-1})|\psi\ket,
\end{equation}
we have
\begin{align}
	\|\psi-\psi_\trunc\|^2 &\stackrel{\eqref{eq:errDelta}}{=} \|\phi_1\|^2
	 =   \big\|(1-\hP_1)|\psi\ket\big\|^2 + \big\|\hP_1|\phi_2\ket\big\|^2 \nonumber\\
	&\, \leq \big\|(1-\hP_1)|\psi\ket\big\|^2 + \|\phi_2\|^2 \nonumber\\
	&\, =   \big\|(1-\hP_1)|\psi\ket\big\|^2 + \big\|(1-\hP_2)|\psi\ket\big\|^2 + \big\|\hP_2|\phi_3\ket\big\|^2 \nonumber\\
	&\,\leq \dotsc \leq \sum_{i=1}^{N-1}\big\|(1-\hP_i)|\psi\ket\big\|^2
	\stackrel{\eqref{eq:err}}{=} \sum_{i=1}^{N-1} \veps_i(M_i).
	\label{eq:errTot}
\end{align}

From the main result stated in Section~\ref{sec:Intro}, this proves upper bounds on the achievable approximation accuracy and required bond dimensions in Eqs.~\eqref{eq:MR-errTot} and \eqref{eq:MR-Mbound} expressed in terms of truncation errors \eqref{eq:MR-err}. For the bond-dimension bound in Eq.~\eqref{eq:MR-Mbound}, we have chosen to distribute errors evenly across bonds ($\veps_i\leq \epsilon$ for all bonds $i$).
For MPS, note that the error bound \eqref{eq:errTot} is $1/2$ of the original bound \eqref{eq:MPS-VCbound} by Verstraete and Cirac \cite{Verstraete2005-5}.

\section{Bounds in terms of R\'{e}nyi entanglement entropies}\label{sec:Renyi}
A central idea in Refs.~\cite{Verstraete2005-5,Schuch2008-100a} on MPS was to relate the truncation error \eqref{eq:MR-err} for bond $i$ to the associated single-branch-cut R\'{e}nyi entanglement entropy
\begin{equation}
	S_\alpha(\rho_i)=\frac{1}{1-\alpha}\ln\sum_\mu\rho_{i,\mu}^\alpha\quad
	(\alpha>0)
\end{equation}
of the normalized target state $|\psi\ket$ for a bipartition into subsystems $\L_i$ and $\R_i$. Here,
\begin{equation}
	\rho_{i,\mu}:=\lambda^2_{i,\mu}
	\quad\text{with}\quad
	\rho_{i,1}\geq\rho_{i,2}\geq\dotsc\geq 0
	\quad\text{and}\quad
	\sum_\mu\rho_{i,\mu}=1
\end{equation}
denote the decreasingly ordered eigenvalues of the corresponding reduced density matrices, and the truncation error \eqref{eq:MR-err} for bond dimension $M_i$ takes the form
$\veps(\rho_i,M_i)=\sum_{\mu>M_i}\rho_{i,\mu}$.
The R\'{e}nyi entropy $S_\alpha$ is Schur-concave \cite{Bhatia1997} such that $S_\alpha(\rho)\leq S_\alpha(\rho')$ if the distribution $\rho$ majorizes $\rho'$, i.e., if $\sum_{\mu=1}^m\rho_\mu\geq \sum_{\mu=1}^m\rho'_\mu$ for all $m$.

\Emph{Lower bounds on error and bond dimensions.}\,---
For a given TTNS bond dimension $M_i$ and truncation error $\delta:=\veps(\rho_i,M_i)$, we can bound $S_\alpha(\rho_i)$ from \emph{above} by considering the set of all distributions $\rho$ that have the truncation error $\veps(\rho,M_i)=\delta$ and finding the element $\rho'$ that is majorized by all others. The distribution $\rho'$ has to be as spread out as possible, i.e., it takes the form \cite{Schuch2008-100a}
\begin{equation}
	\rho'_\mu=\begin{cases}
	       (1-\delta)/M_i&\text{for}\ \mu=1,\dotsc,M_i\ \ \text{and}\\
	       \delta/(D-M_i)&\text{for}\ \mu=M_i+1,\dotsc,D,\\
	      \end{cases}
\end{equation}
where $D$ is the available Hilbert space dimension (the smaller of the dimensions of the Hilbert spaces for subsystems $\L_i$ and $\R_i$). Hence,
\begin{align}\nonumber
	S_\alpha(\rho_i) 
	&\leq S_\alpha(\rho') = \frac{1}{1-\alpha}\ln\left[M_i\left(\frac{1-\delta}{M_i}\right)^\alpha + (D-M_i)\left(\frac{\delta}{D-M_i}\right)^\alpha\right]\\
	&\leq \ln M_i +\frac{\alpha}{1-\alpha} \ln(1-\delta)\qquad\text{if}\quad \alpha>1.
	\label{eq:SboundDelta}
\end{align}
The constraint $\alpha>1$ on the R\'{e}nyi parameter is needed for the second inequality. Inequality \eqref{eq:SboundDelta} establishes the lower bounds on the achievable approximation accuracy and required bond dimensions in terms of R\'{e}nyi entanglement entropies $S_\alpha$ as stated in Eqs.~\eqref{eq:MR-errTot} and \eqref{eq:MR-Mbound} of the main result.

\Emph{Upper bounds on error and bond dimensions.}\,---
Similarly, for a given TTNS bond dimension $M_i$, truncation error $\epsilon:=\veps(\rho_i,M_i)$, and probability $p:=\rho_{i,M_i}$, we can bound $S_\alpha(\rho_i)$ from \emph{below} by considering the set of all distributions $\rho$ that have the truncation error $\veps(\rho,M_i)=\epsilon$ and $\rho_{M_i}=p$ and finding the element $\rhot$ that majorizes all others as discussed in Refs.~\cite{Verstraete2005-5,Barthel2017_08}. The distribution $\rhot$ has to be as concentrated as possible and, hence, takes the form
\begin{equation}
	\rhot_\mu=\begin{cases}
		1-\epsilon-(M_i-1)p	&\text{for}\ \mu=1,\\
		p					&\text{for}\ \mu=2,\dotsc,M_i+K,\\
		\gamma p			&\text{for}\ \mu=M_i+K+1,\quad\text{and}\\
		0					&\text{for}\ \mu>M_i+K+1,
		\end{cases}
\end{equation}
where $K=\lfloor\epsilon/p\rfloor$ and $\gamma=\epsilon/p-K\in[0,1)$. As we are not interested in the $p$ dependence or do not know $p$, we minimize $S_\alpha(\rhot)$ with respect to $p$ to find the optimal distribution for given $\epsilon$ and $M_i$. This is not easily possible in closed form, but we can minimize the lower bound in 
\begin{align}\nonumber
	S_\alpha(\rhot)
	&= \frac{1}{1-\alpha}\ln\left[(\rhot_1)^\alpha + (M_i+K-1+\gamma^\alpha)p^{\alpha}\right]\\
	&\geq \frac{1}{1-\alpha}\ln\left[(M_i+\epsilon/p)p^{\alpha}\right]
	\label{eq:SboundEps-pre}
\end{align}
with respect to $p$, where we have used that $\rhot_1\geq \rhot_2=p$ and assume $\alpha<1$. One finds the minimizer
\begin{equation}
	p^*=\frac{1-\alpha}{\alpha}\frac{\epsilon}{M_i}.
\end{equation}
The condition $\rhot_1(p^*)\geq p^*$ leads to the allowed $\alpha$ interval
\begin{equation}\label{eq:alphaRange}
	\epsilon\leq \alpha < 1.
\end{equation}
Using $p=p^*$ for the lower bound in Eq.~\eqref{eq:SboundEps-pre}, we arrive at the result \cite{Verstraete2005-5,Barthel2017_08}
\begin{equation}\label{eq:SboundEps}
	S_\alpha(\rho_i) \geq S_\alpha(\rhot)
	\geq \frac{1}{1-\alpha}\ln\left[M_i^{1-\alpha}\epsilon^\alpha\frac{(1-\alpha)^{\alpha-1}}{\alpha^\alpha}\right]
	\geq \frac{1}{1-\alpha}\ln\left[M_i^{1-\alpha}\epsilon^\alpha\right].
\end{equation}
In the last step, we have used the non-negativity of the binary entropy function $h(\alpha)=-\alpha\ln\alpha-(1-\alpha)\ln(1-\alpha)$ to simplify the bound.

The inequality~\eqref{eq:SboundEps} can be used to bound the TTNS truncation error $\epsilon$ from above for a given bond dimension $M_i$ or to bound $M_i$ from above for a given desired truncation error $\epsilon$. From the main result stated in Section~\ref{sec:Intro}, this yields the upper bounds on the achievable approximation accuracy and required bond dimensions expressed in terms of R\'{e}nyi entanglement entropies $S_\alpha$ in Eqs.~\eqref{eq:MR-errTot} and \eqref{eq:MR-Mbound}.

If R\'{e}nyi entanglement entropies $S_{\alpha,i}$ are known for several values of the parameter $\alpha$, one can optimize the bounds \eqref{eq:SboundDelta} and \eqref{eq:SboundEps} with respect to $\alpha$.

\section{Conclusion}\label{sec:Conclusion}
The main result stated in the introduction provides rigorous bounds \eqref{eq:MR-errTot} on the attainable TTNS approximation accuracy for given bond dimensions and, conversely, bounds \eqref{eq:MR-Mbound} on TTNS bond dimensions needed to achieve a certain approximation accuracy. The bounds can be formulated in terms of the target state's Schmidt spectra or (less tightly) in terms of $\alpha$-R\'{e}nyi entanglement entropies $S_{\alpha,i}$ with $\alpha<1$ for upper bounds and $\alpha>1$ for lower bounds.

If the target state obeys an \Emph{entanglement area law} or \Emph{log-area law} for a certain tree graph, then the presented bounds imply the existence of efficient faithful TTNS approximations. Specifically, for a certain assignment of physical sites to vertices of a tree graph, the entanglement entropies $S_{\alpha,i}$ for cutting the tree at any edge $i$ should have an upper bound 
\begin{equation}
	S_{\alpha,i} < C\quad\text{(area law)}
	\qquad\text{or}\qquad\
	S_{\alpha,i} < c\,\ln N\quad\text{(log-area law)}
	\quad\forall\ 1\leq i<N
\end{equation}
with some size-independent real constants $C$ and $c>0$, respectively. Then, TTNS approximations exist for any required accuracy, with bond dimensions growing at most polynomially in the system size $N$.

If the target-state entanglement across some edges grows polynomially in the system size, the lower bound in Eq.~\eqref{eq:MR-Mbound} implies that the associated TTNS bond dimensions need to increase exponentially in $N$.

For \Emph{low-energy states of gapped systems} with finite-range interactions on a tree graph, one generally expects an entanglement area law if correlation lengths are sufficiently short compared to the average vertex degree. The situation is somewhat more nuanced in comparison to one-dimensional systems, where the entanglement area law holds independently of the (finite) amplitude of correlation lengths \cite{Hastings2007-08,Brandao2013-9}.
For multiple gapped spin systems on trees with vertex degree $z$, numerical studies in Refs.~\cite{Nagaj2008-77,Kumar2012-85,Li2012-86} report groundstate correlation lengths $\xi$ bounded from above by $\xi_0=1/\ln(z-1)$, which suggests an entanglement area law. Note that the products $(z-1)^\ell e^{-\ell\xi}$ of the volume growth factor and correlation decay, which appear in entanglement bounds \`{a} la Hastings \cite{Hastings2007-08}, may diverge for correlation lengths $\xi>\xi_0$ with increasing distance $\ell$ from the cut tree edge. In Ref.~\cite{Caravan2014-89} volume-law entanglement $S_{\alpha,i}\sim c' N$ was observed in the ground states of gapless fermionic systems on Bethe lattices.

If the target state does \emph{not} obey an entanglement (log-)area law with respect to a tree graph as, e.g., generally expected for ground states of 2D and 3D lattice systems, the derived bounds can still be useful to \Emph{estimate TTNS computation costs} and control bond dimensions.
They imply, for example, that for low-energy states of typical systems in $D\geq 2$ spatial dimensions, MPS simulations are asymptotically more efficient than TTNS simulations. As argued in Ref.~\cite{Barthel2026_01} based on the bounds~\eqref{eq:MR-Mbound}, the computation cost difference grows exponentially with system size. This conclusion stands in some tension with common expectations in the community and promising numerical TTNS simulations \cite{Tagliacozzo2009-80,Murg2010-82,Kloss2020-9,Cataldi2021-5,Felser2021-126,Krinitsin2025-112,Pavesic2025_09}, but the discrepancy may simply stem from size-dependent polynomial cost factors. The latter are disregarded in Ref.~\cite{Barthel2026_01} and may lead to a crossover system size below which TTNS prevail over MPS.
One can also employ the derived bounds~\eqref{eq:MR-Mbound} to \Emph{optimize the tree graph structure} \cite{Nakatani2013-138,Murg2015-11,Larsson2019-151,Cataldi2021-5,Ferrari2022-105,Hikihara2023-5} to reduce TTNS computation costs, e.g., for applications in quantum chemistry and quantum computation.

While the bounds on approximation accuracy and required bond dimensions can guarantee the existence of efficient TTNS approximations, we do not necessarily have \Emph{efficient methods} for finding these approximations. For the variational optimization of groundstate approximations, DMRG algorithms \cite{Otsuka1996-53,Lepetit2000-13,Murg2010-82,Nakatani2013-138,Cao2021-104} generally work well and we do not expect that more practical methods exist for unconstrained TTNS, but there is no fundamental theory for the convergence of DMRG. Concerning MPS groundstate search for one-dimensional gapped local Hamiltonians, Landau, Vazirani, and Vidick devised a provably polynomial-time algorithm \cite{Landau2015-11}. It is conceivable that this can be extended to TTNS groundstate search for local Hamiltonians on trees with a sufficiently large energy gap.

Finally, as the presented bounds are based on a truncation of the exact target state, they apply to the case of unconstrained TTNS tensors and would generally need adaptation for \Emph{TTNS with tensor constraints} such as tensor homogeneity and scale invariance which drastically reduce the number of degrees of freedom \cite{Martin-Delgado1996-76,Vidal-2005-12,Nagaj2008-77,Giovannetti2008-101,Giovannetti2009-79,Pfeifer2009-79,Tagliacozzo2009-80,Nagy2012-327,Li2012-86}, restrictions on CP ranks which can drastically lower contraction costs \cite{Chen2022_05,Chen2024-46}, and imposition of substructures like brickwork circuits of two-qubit gates (a.k.a.\ tensor Trotterization) which is useful for implementations on quantum computers \cite{Miao2021_08,Miao2023_03,Haghshenas2022-12,Haghshenas2023_05} and can lower classical contraction costs \cite{Barthel2025-111}. The \emph{lower} bounds on the approximation accuracy and required bond dimension in Eqs.~\eqref{eq:MR-errTot} and \eqref{eq:MR-Mbound} of course still apply.

\begin{acknowledgments}
I thank Elizabeth Bennewitz, Zohreh Davoudi, Alexey Gorshkov, So-Yun Jang, Hersh Kumar, Alessio Lerose, Benjamin Lovitz, Federica Surace, and Nikita Zemlevskiy for discussions that motivated this work as well as Simone Montangero, Luka Pave\v{s}i\'{c}, and Darvin Wanisch for helpful feedback.
\end{acknowledgments}

\end{document}